\documentclass[a4paper]{jpconf}

\usepackage{bm}
\usepackage{latexsym}
\usepackage{amsfonts,amssymb,amsmath}
\usepackage{graphicx,epsfig}
\usepackage{psfrag}
\usepackage{subfigure}
\usepackage{ulem}
\usepackage{hyperref}
\usepackage{color}
\usepackage{appendix}

\newcommand{\dd}{\mathrm{d}}

\def\spose#1{\hbox to 0pt{#1\hss}}

\def\lta{\mathrel{\spose{\lower 3pt\hbox{$\mathchar"218$}}
     \raise 2.0pt\hbox{$\mathchar"13C$}}}
\def\gta{\mathrel{\spose{\lower 3pt\hbox{$\mathchar"218$}}
     \raise 2.0pt\hbox{$\mathchar"13E$}}}

\newcommand{\ie}{\textsl{i.e.~}}

\newcommand{\eg}{\textsl{e.g.~}}

\newcommand{\Hu}{\mathcal{H}}

\newcommand{\Mp}{M_{_\mathrm{Pl}}}

\def\beq{\begin{equation}}
\def\eeq{\end{equation}}
\def\bea{\begin{eqnarray}}
\def\eea{\end{eqnarray}}
\def\eqref{\ref}

\newcommand{\Rea}{\Re \mathrm{e}\,}
\newcommand{\Ima}{\Im \mathrm{m}\,}

\begin{document}

\title{The Quantum State of Inflationary Perturbations}

\author{J\'er\^ome Martin}

\address{Institut d'Astrophysique de Paris, UMR 7095-CNRS,
Universit\'e Pierre et Marie Curie, 98 bis boulevard Arago, 75014
Paris, France}

\ead{jmartin@iap.fr}

\begin{abstract}
  This article reviews the properties of the quantum state of
  inflationary perturbations. After a brief description of the
  inflationary background, the wavefunction of the Mukhanov-Sasaki
  variable is calculated and shown to be that of a strongly squeezed
  state. The corresponding Wigner function and density matrix, which
  are convenient tools to characterize the properties of a quantum
  state, are also evaluated. Finally, the issue of definite outcomes
  for inflation is discussed.
\end{abstract}

\section{Introduction}
\label{sec:intro}

According to the theory of inflation, the very early universe
underwent a period of quasi-exponential
expansion~\cite{Starobinsky:1980te,Guth:1980zm,Linde:1981mu,Albrecht:1982wi,Linde:1983gd}. Such
a phase of evolution can arise in general relativity when the pressure
of the fluid sourcing the Einstein equations is negative. This is the
case if a scalar field with a flat potential dominated in the very
early universe. This simple assumption, which solely rests on
classical physics, is quite remarkable as it allows us to solve many
puzzles of the former standard model of cosmology, \ie the hot Big
Bang scenario.

\par

But even more interesting results can be obtained when quantum effects
are taken into account. In particular, considering small inhomogeneous
fluctuations on top of an isotropic and homogeneous expanding
background and quantizing them gives a very elegant explanation for
the origin of the Cosmic Microwave Background (CMB) anisotropies and
of the large scale
structures~\cite{Mukhanov:1981xt,Mukhanov:1982nu,Starobinsky:1982ee,Guth:1982ec,Hawking:1982cz,Bardeen:1983qw}
(for reviews, see
Refs.~\cite{Mukhanov:1990me,Martin:2003bt,Martin:2004um,Martin:2007bw,Linde:2007fr,Sriramkumar:2009kg,Peter:2009}). As
is well-known, this leads to an almost scale-invariant power spectrum
for the cosmological perturbations~\cite{Stewart:1993bc}, the small
deviations from scale invariance being connected to the detailed
shape of the scalar field potential. As a consequence, this provides a
method to constrain the physical conditions that prevailed when the
universe was very young, at energy scales that can be as high as
$10^{15}
\mbox{GeV}$~\cite{Larson:2010gs,Komatsu:2010fb,Martin:2006rs,Lorenz:2007ze,Lorenz:2008je,Martin:2010kz,Martin:2010hh}.

\par

This is not, however, the only interesting aspect that arises when
quantum effects are included into the inflationary picture. Very
fundamental questions, connected to the quantum nature of the
gravitational field and/or to the foundations of quantum mechanics,
can also be discussed in this framework. In these proceedings, we
review the properties of the quantum state of inflationary
perturbations \cite{Grishchuk:1990bj,Grishchuk:1992tw} and briefly
discuss how it is related to the more fundamental issues mentioned
above~\cite{Grishchuk:1990bj,Grishchuk:1992tw,Polarski:1995jg,Lesgourgues:1996jc,Grishchuk:1997pk,Kiefer:2006je,Kiefer:1998qe,Egusquiza:1997ez,Anderson:2005hi,Burgess:2006jn,Martineau:2006ki,Sudarsky:2009za}. This
short article is organized as follows. In the next section,
Sec.~\ref{sec:back}, we describe inflation at the background
level. Then, in Sec.~\ref{sec:pert}, we review the theory of
inflationary fluctuations of quantum-mechanical origin, paying special
attention to the quantum state in which the perturbations are
placed. Finally, in Sec.~\ref{sec:conclusions}, we present our
conclusions and briefly mention and discuss the open issues.

\section{The Inflationary Background}
\label{sec:back}

Inflation is a phase of accelerated, quasi-exponential, expansion that
took place in the very early Universe, prior to the standard hot
Big-Bang phase
~\cite{Starobinsky:1980te,Guth:1980zm,Linde:1981mu,Albrecht:1982wi,Linde:1983gd}
(for reviews, see
Refs.~\cite{Martin:2003bt,Martin:2004um,Martin:2007bw}). It is
important to realize that inflation does not replace the former
cosmological standard model by a new one but just completes
it. Postulating such a phase of evolution allows us to solve the
standard problems of the hot Big-Bang model. Given that, at very high
energies, field theory is the relevant framework to describe matter, a
natural way to realize inflation is to consider that a real scalar
field (the ``inflaton'' field) was the dominant source of energy
density in the early Universe (there also exist versions of the
inflationary scenario where inflation is driven by several scalar
fields). This assumption is moreover compatible with the observed
homogeneity, isotropy and flatness of the early Universe. Technically,
this can be described by the metric tensor ${\rm d}s^2=-{\rm
  d}t^2+a^2(t)\delta _{ij}{\rm d}x^i{\rm d}x^j$, where $a(t)$ is the
Friedman-Lema\^{\i}tre-Robertson-Walker (FLRW) scale factor and $t$
the cosmic time. The Einstein equations imply that
$\ddot{a}/a=-(\rho+3p)/(6\Mp^2)$, $\rho$ and $p$ being the energy
density and pressure of the matter sourcing the gravitational field
and $\Mp$ the Planck mass (a dot denotes a derivative with respect to
the cosmic time $t$). For a scalar field, this reduces to
$\ddot{a}/a=V(\varphi)(1-\dot{\varphi}^2/V)/(3\Mp^2)$, where
$V(\varphi)$ is the scalar field potential. This means that inflation
(\ie $\ddot{a}>0$) can be obtained provided the inflaton slowly rolls
down its potential so that its potential energy dominates over its
kinetic energy. This also shows that the inflaton potential must be
sufficiently flat, a requirement which is not always easy to obtain in
realistic theories and makes the inflationary model building problem a
difficult issue \cite{Lyth:1998xn}. The physical nature of the
inflaton field has not been identified (there are many candidates)
and, as a consequence, the shape of $V(\varphi)$ is not known. Of
course, different $V(\varphi)$ lead to different inflationary
expansions but, since these different potentials must all be
sufficiently flat, the corresponding scale factors are all
approximately given by the de Sitter solution. This solution is
described by the scale factor $a(t)\simeq {\rm e}^{Ht}$, where
$H\equiv \dot{a}/a$ is the Hubble parameter, a slowly-varying quantity
directly related to the energy scale of inflation. Observationally,
this latter quantity is not known but is constrained
~\cite{Martin:2006rs} to be between the Grand Unified Theory (GUT)
scale, that is to say $\sim 10^{15}\, \mbox{GeV}$, and $\sim 1\,
\mbox{TeV}$. It is also interesting to remark that the energy density
during inflation, $\rho_{\rm inf}\simeq 3H^2_{\rm inf}\Mp^2$, remains
approximately constant. This implies the remarkable property that the
classical background approximation remains valid even if one runs the
clock backwards and study the initial stages of inflation. This would
no longer be the case if an ordinary fluid (\eg radiation or matter)
dominated the energy budget of the early Universe since the scaling
$\rho \propto 1/a^4$ or $\rho \propto 1/a^3$ would quickly imply that
$\rho \gg \Mp^4$ at early times. Of course, this does not mean that
quantum effects play no role during inflation. As a matter of fact,
when the quantum kicks $\Delta \varphi_{\rm q}$ are much larger than
the typical classical motion $\Delta \varphi_{\rm cl}\simeq
V'/(3H^2)$, they must be taken into account. This can be done in the
framework of stochastic
inflation~\cite{Starobinsky:1986fx,Martin:2005ir}.

\par

But the arguably most impressive result coming from including the
quantum effects into the inflationary picture concerns the origin of
the cosmological fluctuations that give rise to the Cosmic Microwave
Background (CMB) anisotropies and to the large-scale structures. We
turn to this question in the next section.

\section{Inflationary Perturbations}
\label{sec:pert}

According to inflation, the seeds of cosmological fluctuations are the
quantum fluctuations of the inflaton and gravitational fields. As is
well-known, this idea leads to predictions that are in remarkable
agreement with astrophysical
data~\cite{Larson:2010gs,Komatsu:2010fb,Martin:2006rs,Lorenz:2007ze,Lorenz:2008je,Martin:2010kz,Martin:2010hh}. From
the theoretical point of view, the perturbations are viewed as small
ripples on top of a classical background in close analogy with the
condensed matter situation where phonons are considered to be
quantized excitations propagating within a classical crystal. In order
to describe cosmological perturbations, one has to go beyond the
homogeneity and isotropy of the FRLW metric. This can be accomplished
by writing the metric as~\cite{Mukhanov:1990me} $ \dd
s^2=a^2\left(\eta\right)\Bigl\lbrace -\left(1-2\phi\right)\dd\eta^2
+2\left(\partial_iB\right)\dd x^i\dd \eta
+\left[\left(1-2\psi\right)\delta_{ij} +2\partial_i\partial_jE
\right]\dd x^i\dd x^j\Bigr\rbrace $, where $\eta$ is the conformal
time and where the four functions $\phi$, $B$, $\psi$ and $E$ are
functions of time and space. The above approach is, however, redundant
because of gauge freedom
\cite{Mukhanov:1990me,Bardeen:1980kt,Martin:1997zd}. The gravitational
sector can in fact be described by a single, gauge-invariant,
quantity, the so-called Bardeen potential $\Phi_{_{\rm B}}$ defined by
\cite{Bardeen:1980kt} $ \Phi_{_{\rm B}}\left(\eta,\bm{x}\right)= \phi+
\left[a\left(B-E^\prime\right)\right]^\prime/a, $ where a prime
denotes a derivative with respect to the conformal time $\eta$.  In
the same manner, the matter sector can be modeled by the gauge
invariant fluctuation of the scalar field $
\delta\varphi^{\left(\mathrm{gi}\right)}\left(\eta,\bm{x}\right)=
\delta\varphi+\varphi^\prime\left(B-E^\prime\right)$. The two
quantities $\Phi_{_{\rm B}}$ and
$\delta\varphi^{\left(\mathrm{gi}\right)}$ are related by a perturbed
Einstein constraint. This implies that the scalar sector can in fact
be described by a single quantity. This quantity is the
Mukhanov-Sasaki variable \cite{Mukhanov:1981xt,Kodama:1985bj} which is
a combination of the Bardeen potential and of the gauge invariant
field, namely $v\left(\eta,\bm{x}\right)=a
\left[\delta\varphi^{\left(\mathrm{gi}\right)}
  +\varphi^\prime\Phi_{_{\rm B}}/\Hu\right]$, where ${\cal H}\equiv
a'/a$. It is also convenient to Fourier transform $v(\eta,{\bm x})$
since we consider a linear theory and, hence, all the modes evolve
independently. Therefore, instead of $v(\eta,\bm{x})$, we will rather
work with its Fourier amplitude $v_{\bm{k}}(\eta)$. Its equation of
motion can be easily obtained from the perturbed Einstein equations
and reads
\begin{equation}
\label{eq:eomv}
v_{\bm k}''+\omega^2\left(\eta, \bm{k}\right)v_{\bm k}=0,
\end{equation}
where the time-dependent frequency $\omega(\eta, \bm{k})$ can be
written as $\omega^2\left(\eta, \bm{k}\right)=k^2-
\left(a\sqrt{\epsilon_1}\right)^{\prime\prime}/(a\sqrt{\epsilon_1})$,
$k$ being the wavenumber of the mode under consideration and
$\epsilon_1\equiv -\dot{H}/H^2$ the first slow-roll parameter
characterizing the cosmological expansion during inflation. The above
equation is the equation of a parametric oscillator (\ie similar to a
pendulum the length of which can change with time). In the Minkowski
situation, the scale factor is a constant and $\omega=k$. In this
case, one deals with an harmonic oscillator. We see that the time
dependence of the frequency $\omega(\eta ,\bm{k})$ is entirely due to
the fact that the perturbations live in a dynamical spacetime. It is
worth noticing that Eq.~(\ref{eq:eomv}) appears in other interesting
physical situations such as the Schwinger
effect~\cite{Schwinger:1951nm,Brezin:1970xf} or the dynamical Casimir
effect~\cite{1970JMP....11.2679M} (recently observed for the first
time in the laboratory, see Ref.~\cite{2011Natur.479..376W}) and, each
time, the same phenomenon is observed: particle creation under the
influence of the classical, time-dependent, source even if the details
of the corresponding process (number of pairs created, power spectrum
etc ...) depend on the precise time evolution of the source. In this
sense, the theory of cosmological perturbations can be viewed as
``standard'' in the framework of quantum field theory. The only new
aspect is that the time-dependent source is the classical
gravitational background of the expanding space-time. Let us also
notice that inflation is in fact not necessary in order to have
particle creation: as is clear from the above considerations, only a
time-dependent scale factor is necessary. However, the inflationary
expansion becomes mandatory if one wants a $\omega(\eta, \bm{k})$
which leads to a power spectrum in agreement with the CMB
measurements.

\par

\begin{figure*}[t]
\begin{center}
\includegraphics[height=7cm,width=14.cm]{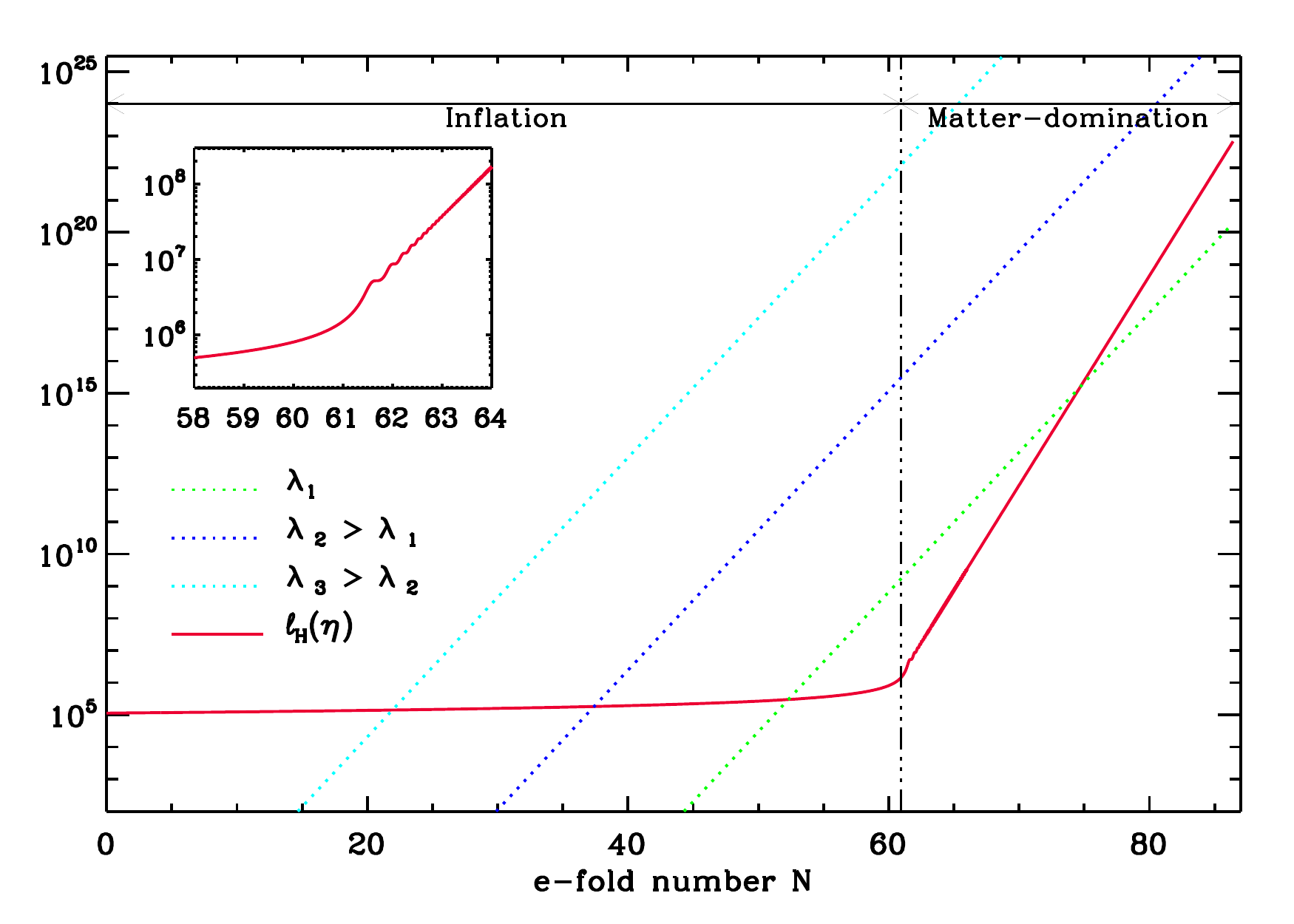}
\end{center}
\caption{Evolution of various relevant length scales (in Planck units)
  during and after inflation. The solid red line represents the Hubble
  radius $\ell_{_{\rm H}}\equiv 1/H$. It is almost constant during
  inflation and starts to grow during the post-inflation era (during
  this phase, the inflaton field oscillates at the bottom of its
  potential and the universe becomes matter-dominated. Small
  oscillations in $\ell_{_{\rm H}}$ are clearly visible in the
  inset). The three dashed lines represent the physical wavelengths of
  three different Fourier modes. Initially within the Hubble radius,
  they exit $\ell_{_{\rm H}}$ during inflation and re-enter it in the
  post-inflationary era.  }
\label{fig:scaleinf}
\end{figure*}

We now describe how the cosmological fluctuations are quantized. In
order to achieve this goal, it is also more convenient to work with
real variables. Therefore, we introduce the following definitions
$v_{\bm{k}}\equiv \left(v_{\bm{k}}^\mathrm{{\rm R}}+
  iv_{\bm{k}}^\mathrm{{\rm I}}\right)/\sqrt{2}$ and for the conjugate
momentum of the Mukhanov-Sasaki variable, $p_{\bm{k}}\equiv
\left(p_{\bm{k}}^\mathrm{{\rm R}}+ ip_{\bm{k}}^\mathrm{{\rm
      I}}\right)/\sqrt{2}$. In the Schr\"odinger approach, the quantum
state of the system is described by a wavefunctional,
$\Psi\left[v(\eta,{\bm x})\right]$. Since we are working in Fourier
space (and since the theory is still free in the sense that it does
not contain terms with power higher than two in the Lagrangian), the
wavefunctional can also be factorized into mode components as
\begin{equation}
\Psi\left[v(\eta,{\bm x})\right]=\prod _{\bm k}
\Psi_{\bm k}\left(v_{\bm{k}}^\mathrm{R},
v_{\bm{k}}^\mathrm{I}\right)
=\prod_{\bm k}\Psi^{\rm R}_{\bm k}\left(v_{\bm{k}}^\mathrm{R}\right)
\Psi ^{\rm I}_{\bm k}\left(v_{\bm{k}}^\mathrm{I}\right).
\end{equation}
It can be shown that the solution to the Schr\"odinger equation can be
written as a Gaussian function, which should not come as a surprise
since we are quantizing an oscillator. Concretely, one can write
\begin{equation}
\label{eq:gaussianpsi}
\Psi_{\bm k}^{\rm R,I}\left(\eta,v_{\bm k}^{\rm R,I}\right)
=N_{\bm k}(\eta){\rm e}^{-\Omega_{\bm k}(\eta)\left(v_{\bm k}^{\rm R,I}\right)^2}.
\end{equation}
where the functions $N_{\bm k}(\eta)$ and $\Omega _{\bm k}(\eta)$
(which do not depend on whether one considers the wavefunction of the
real or imaginary parts of the Mukhanov-Sasaki variable) can be
expressed as
\begin{equation}
\label{eq:solpsi}
\left \vert N_{\bm k}\right \vert 
=\left(\frac{2\Rea  \Omega _{\bm k}}{\pi}\right)^{1/4}, \quad 
\Omega_{\bm k}=-\frac{i}{2}\frac{f_{\bm k}'}{f_{\bm k}},
\end{equation}
where $f_{\bm k}$ is a function obeying the equation $f_{\bm
  k}''+\omega^2f_{\bm k}=0$, that is to say exactly
Eq.~(\ref{eq:eomv}). 

\par

We now discuss the question of the initial conditions. The fundamental
assumption of inflation is that the perturbations are initially in
their ground state. At the beginning of inflation, all the modes of
astrophysical interest today have a physical wavelength smaller than
the Hubble radius, see Fig.~\ref{fig:scaleinf}. In this regime, each
mode now behaves as a harmonic oscillator (as opposed to a parametric
oscillator in the generic case) with frequency $\omega=k$. As a
consequence, the differential equation for $f_{\bm k}(\eta)$ can
easily be solved and the solution reads $f_{\bm k}=A_{\bm k}{\rm
  e}^{ik\eta}+B_{\bm k}{\rm e}^{-ik\eta}$, $A_{\bm k}$ and $B_{\bm k}$
being integration constants. Upon using the second
equation~(\ref{eq:solpsi}), one has
\begin{equation}
\Omega_{\bm k}\rightarrow \frac{k}{2}\frac{A_{\bm
  k}{\rm e}^{ik\eta}-B_{\bm k}{\rm e}^{-ik\eta}}{A_{\bm k}{\rm e}^{ik\eta}
+B_{\bm k}{\rm e}^{-ik\eta}}.
\end{equation}
The wavefunction~(\ref{eq:gaussianpsi}) represents the ground state
wavefunction of a harmonic oscillator if $\Omega_{\bm
  k}=k/2$. Therefore, one must choose the initial conditions such that
$B_{\bm k}=0$, the value of $A_{\bm k}$ turning out to be irrelevant
in this regime. This choice of initial conditions completely
determines the solution and is referred to as the Bunch-Davies
vacuum~\cite{Birrell:1982ix}.

\par

We have just seen that, when the wavelength of a Fourier mode is
smaller than the Hubble radius, the corresponding wavefunction can be
chosen to be the ground state wavefunction of a harmonic
oscillator. However, this is no longer the case when the mode exits
the Hubble radius, see Fig.~\ref{fig:scaleinf}. In this regime, the
function $\Omega_{\bm{k}}(\eta)$ no longer equals $k/2$ and acquires a
non trivial time-dependence. In fact, a time dependent wavefunction as
the one in Eq.~(\ref{eq:gaussianpsi}) represents a very peculiar
quantum state known as a squeezed state. A squeezed state is a
Gaussian state for which there exists a direction in the plane
$(p_{\bm{k}},v_{\bm{k}})$ where the dispersion is extremely small, \ie
exponentially small $\simeq {\rm e}^{-2r_{\bm{k}}}$ where $r_{\bm{k}}$
is the squeezing parameter (of course, the dispersion in the
perpendicular direction is very large $\simeq {\rm e}^{2r_{\bm{k}}}$
in order to satisfy the Heisenberg relation). The phenomenon of
squeezing is widely studied in many different branches of physics, in
particular in quantum optics. Squeezing occurs each time the
quantization of a parametric oscillator is carried out. It is
remarkable that the quantization of small fluctuations on top of an
expanding universe also leads to that phenomenon. Another interesting
feature is that cosmological squeezing is much larger than what can be
achieved in the laboratory~\cite{2008PhRvL.100c3602V}: for modes of
cosmological interest today, $r_{\bm{k}}$ can reach values of order
$\sim 10^2$~\cite{Grishchuk:1990bj}. In the literature, this regime is
very often described as a regime where the cosmological perturbations
have
classicalized~\cite{Lesgourgues:1996jc,Polarski:1995jg,Kiefer:1998qe,Kiefer:1998qe,Kiefer:1999gt}. Since
this concept is subtle in quantum mechanics, we need to come back to
this issue and to describe accurately what is meant by a ``classical
limit'' in this context. In particular, it may seem strange at first
sight that a quantum system placed in a strongly squeezed state can be
described as a classical state since, in the context of, say, quantum
optics, a similar situation would precisely be described as a
non-classical situation~\cite{Caves:1985zz,Schumaker:1985zz}.

\par

\begin{figure*}[t]
\begin{center}
\includegraphics[width=3.5cm]{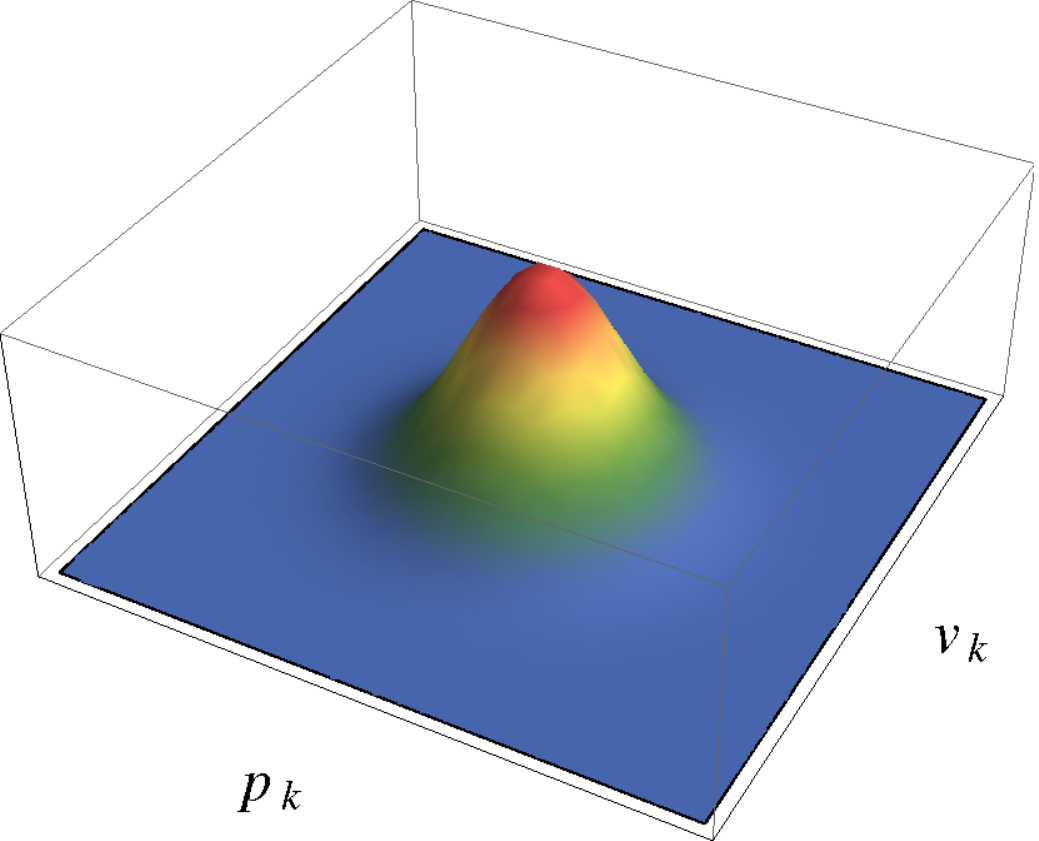}
\includegraphics[width=3.5cm]{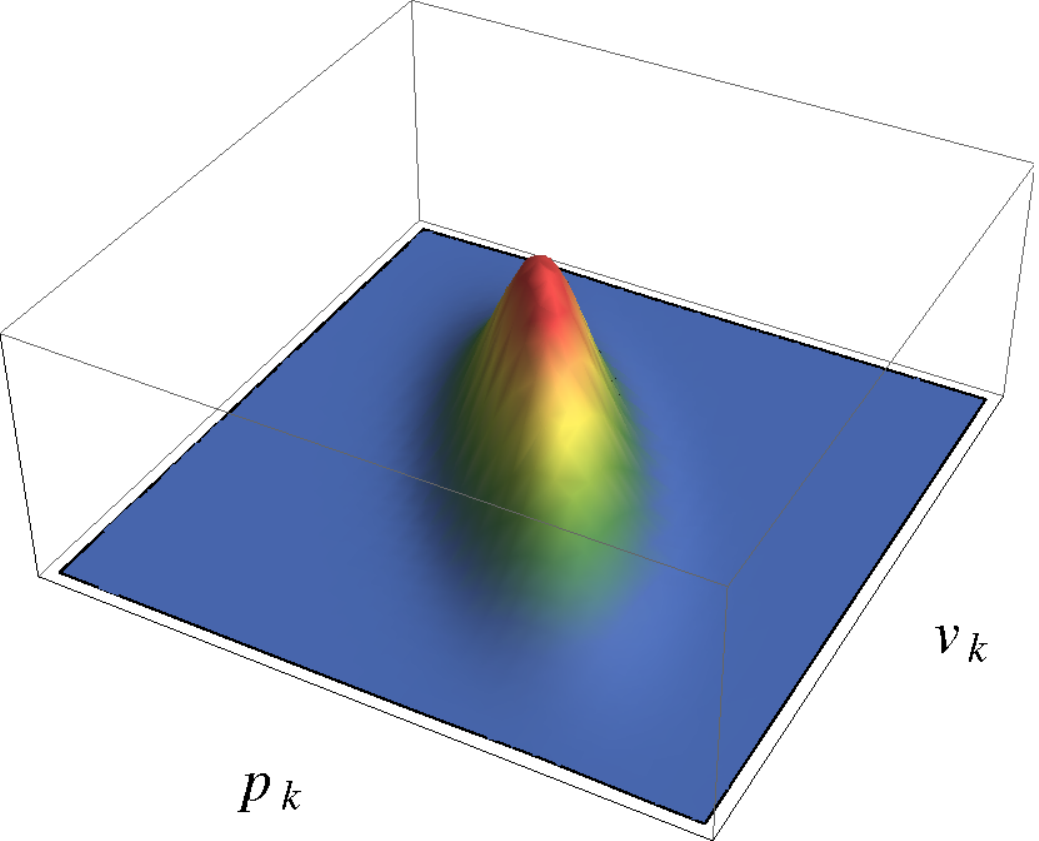}
\includegraphics[width=3.5cm]{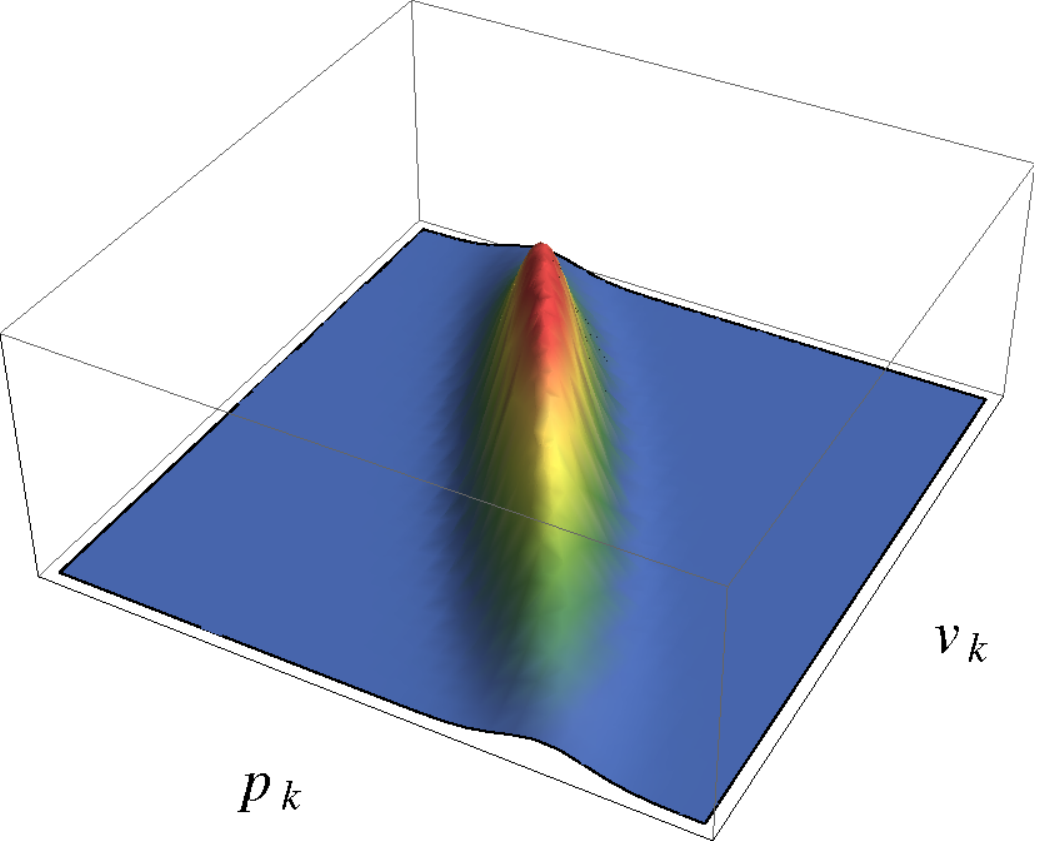}
\includegraphics[width=3.5cm]{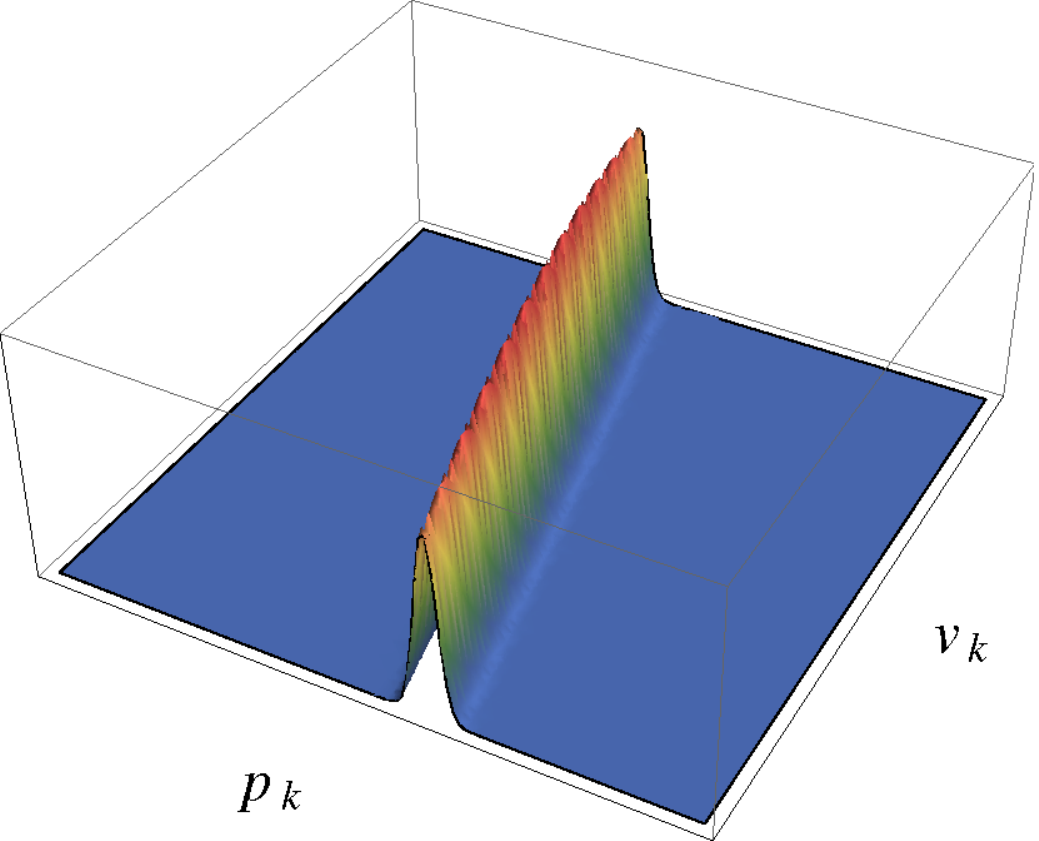}
\end{center}
\caption{Wigner function of a squeezed quantum state at different
  times during inflation. Only the two-dimensional function
  corresponding to the set of variables $\left(v_{\bm k}^{\rm
      R},p_{\bm k}^{\rm R}\right)$ has been represented, see
  Eq.~(\ref{eq:wignersqueeze}). The time evolution of $\Rea \Omega
  _{\bm k}$ and $\Ima \Omega _{\bm k}$ has been calculated using the
  Schr\"odinger equation with the Hamiltonian that can be obtained
  from the relativistic theory of cosmological perturbations. The left
  panel corresponds to $r_{\bm k}=0.0005$ and the corresponding state
  is almost a coherent one. Then, from left to right, one has $r_{\bm
    k}=0.48$, $r_{\bm k}=0.88$ and, finally, $r_{\bm k}=2.31$ for the
  right panel. The effect of the squeezing and the cigar shape of
  Eq.~(\ref{eq:wignersqueeze}) are clearly visible.}
\label{fig:wigner}
\end{figure*}

A convenient tool to study this question is the Wigner function, defined by
\begin{equation}
W\left(v_{\bm k}^{\rm R},v_{\bm k}^{\rm I},p_{\bm k}^{\rm R},p_{\bm k}^{\rm I}\right)
= \frac{1}{(2\pi)^2}
\int {\rm d}x{\rm d}y \, \Psi^*\left(v_{\bm k}^{\rm R}-\frac{x}{2}, 
v_{\bm k}^{\rm I}-\frac{y}{2}\right)
{\rm e}^{-ip_{\bm k}^{\rm R}x-ip_{\bm k}^{\rm I}y}\, 
\Psi\left(v_{\bm k}^{\rm R}+\frac{x}{2}, 
v_{\bm k}^{\rm I}+\frac{y}{2}\right).
\end{equation}
It is well known that the Wigner function can be understood as a
classical probability distribution function whenever it is positive
definite. Then, upon using the quantum state~(\ref{eq:gaussianpsi}),
the following explicit form is obtained
\begin{align}
\label{eq:wignersqueeze}
W\left(v_{\bm k}^{\rm R},v_{\bm k}^{\rm I},p_{\bm k}^{\rm R},p_{\bm k}^{\rm I}\right)
=\Psi\Psi^*\frac{1}{2\pi \Rea  \Omega _{\bm k}} &
\exp\left[-\frac{1}{2\Rea  \Omega _{\bm k}}\left(p_{\bm k}^{\rm R}+2\Ima  \Omega _{\bm k}
v_{\bm k}^{\rm R}\right)^2\right] \nonumber \\ &\times 
\exp\left[-\frac{1}{2\Rea  \Omega _{\bm k}}\left(p_{\bm k}^{\rm I}+2\Ima  \Omega _{\bm k}
v_{\bm k}^{\rm I}\right)^2\right].
\end{align}
This Wigner function~(\ref{eq:wignersqueeze}) is represented in
Fig.~\ref{fig:wigner} at different times or, equivalently, at
different values of $r_{\bm k}$ ($r_{\bm k}=0.0005$, $0.48$, $0.88$
and $2.31$). The effect of strong squeezing is clearly
visible. Initially, in the sub-Hubble regime, $r_{\bm k}$ is small and
the Wigner function is peaked over of small region in phase space. In
this regime, $W$ is just the Wigner function of a coherent state since
we chose to start from the ground state of a harmonic
oscillator. Coherent states are considered as the ``most classical''
states precisely for this reason: if one is given, say, the value of
$v_{\bm k}^{\rm R}$, then one obtains a value for the momentum,
$p_{\bm k}^{\rm R}$, which is very close to the one one would have
inferred in the classical case. This is of course due to the fact that
the Wigner function follows the classical trajectory and has minimal
spread around it in all phase space directions.

\par

As inflation proceeds, the situation described before changes. The
modes become super Hubble and $r_{\bm k}(\eta)$ increases. As a
consequence, the Wigner function spreads and acquires a cigar shape
typical of squeezed states~\cite{Grishchuk:1990bj}. Therefore, if one
is now given $p_{\bm k}^{\rm R}$ then the value of $v_{\bm k}^{\rm R}$
is very uncertain since the Wigner function is spread over a large
region in phase space. In this case, cosmological perturbations
certainly do not behave classically in the usual sense. Given the
previous discussion, it may seem relatively easy to observe genuine
quantum effects in the CMB. Unfortunately this is not so, essentially
because, in the strong squeezed limit, all quantum predictions can be
in fact obtained from averages performed by mean of a classical
stochastic process. The point is that, in the limit $r_{\bm
  k}\rightarrow \infty$, all the quantum predictions can be reproduced
if one assumes that the system always followed classical laws but had
random initial conditions with a given probability density function.
This is possible because the Wigner function, despite being
delocalized over phase space, remains positive. This property is not
general but is a key feature of Gaussian states, hence of squeezed
states.

\par

Let us say a few words about the density matrix $\hat{\rho}_{\bm
  k}^{\rm R}$. The density matrix is nothing but the Fourier transform
of the Wigner function. Let us denote by $\vert v_{\bm k}^{\rm
  R}\rangle$ the eigenstates of the operator $\hat{v}_{\bm k}^{\rm
  R}$. Then, we have
\begin{equation}
\left\langle v_{\bm k}^{\rm R}{}'\left \vert \hat{\rho}_{\bm k}^{\rm R}
\right \vert  v_{\bm k}^{\rm R}\right\rangle =\int _{-\infty}^{\infty}
  {\rm d}y\, {\rm e}^{iy(v_{\bm
      k}^{\rm R}{}'-v_{\bm
      k}^{\rm R})}W\left(\frac{v_{\bm
        k}^{\rm R}{}'+v_{\bm
        k}^{\rm R}}{2},y\right).
\end{equation}
Upon using Eq.~(\ref{eq:wignersqueeze}) in the above equation, one
obtains the following expression
\begin{eqnarray}
\left\langle v_{\bm k}^{\rm R}{}'\left \vert \hat{\rho}_{\bm k}^{\rm R}
\right \vert  v_{\bm k}^{\rm R}\right\rangle &=&
\left(\frac{2\Rea  \Omega _{\bm k}}{\pi}
\right)^{1/2}
{\rm e}^{-\Rea  \Omega _{\bm k}\left[\left(v_{\bm
        k}^{\rm R}{}'\right)^2+\left(v_{\bm
        k}^{\rm R}\right)^2\right]}
{\rm e}^{-i\Ima  \Omega _{\bm k}\left[\left(v_{\bm
        k}^{\rm R}{}'\right)^2-\left(v_{\bm
        k}^{\rm R}\right)^2\right]}\, .
\end{eqnarray}
We notice that the off-diagonal terms, $v_{\bm k}^{\rm R}{}' \neq
v_{\bm k}^{\rm R}$, oscillate very rapidly in the strong squeezing
limit where $\Ima \Omega_{\bm k}\rightarrow \infty$. This means that
decoherence (defined as the disappearance of those off-diagonal terms)
does not occur without taking into account the influence of an
environment for the perturbations. Various discussions on what this
environment may be can be found in
Refs.~\cite{Anderson:2005hi,Burgess:2006jn,Martineau:2006ki}. 

\par

We end this section by a brief discussion of how the theory presented
above can make astrophysical predictions. The perturbations described
by the Mukhanov-Sasaki variable are directly related to CMB
fluctuations. As a consequence, in order to calculate the statistical
properties of the CMB, it is sufficient to evaluate the power spectrum
of the operator $\hat{v}(\eta ,{\bm x})$ (and the higher order
correlation functions if one is interested in non-Gaussianity). It is
given by the following expression
\begin{equation}
\left \langle \Psi\left \vert \hat{v}(\eta ,{\bm x})
\hat{v}(\eta,{\bm x}+{\bm r})\right \vert \Psi\right \rangle
=\int \prod _{\bm k}{\rm d}v_{\bm k}^{\rm R}{\rm d}v_{\bm k}^{\rm I}
\Psi_{\bm k}^*(v_{\bm k}^{\rm R},v_{\bm k}^{\rm I})
v(\eta ,{\bm x})
v(\eta,{\bm x}+{\bm r})
\Psi_{\bm k}(v_{\bm k}^{\rm R},v_{\bm k}^{\rm I}).
\end{equation}
and, upon using the results exposed before, can be expressed as
\begin{eqnarray}
\left \langle \Psi\left \vert \hat{v}(\eta ,{\bm x})
\hat{v}(\eta,{\bm x}+{\bm r})\right \vert \Psi\right \rangle
=\int_0^{+\infty}
\frac{{\rm d}k}{k}\frac{\sin kr}{kr}{\cal P}_v(k)\, ,
\end{eqnarray}
which, in fact, defines the power spectrum ${\cal P}_v(k)$ as the
square of the Fourier amplitude per logarithmic interval at a given
scale, \ie ${\cal P}_v(k)=k^3\vert f_{\bm k}\vert^2/(2\pi^2)$.  On
large scales, one obtains ${\cal P}_{\zeta}(k)\simeq A_{_{\rm
    S}}k^{n_{_{\rm S}}-1}$ where the coefficients $A_{_{\rm S}}$ and
$n_{_{\rm S}}$ (the spectral index) depend on the inflationary model
under consideration. For any ``good'' model of inflation, one has
$n_{_{\rm S}}\simeq 1$ but $n_{_{\rm S}}\neq 1$. This means that
inflation does not predict a scale-invariant power spectrum as
sometimes said but an almost scale-invariant power spectrum. This
difference is not a detail as a scale-invariant power spectrum (or a
Harrison-Zeldovitch power spectrum) was considered well before the
advent of inflation. In this case, $n_{_{\rm S}}=1$ would just be a
postdiction. On the contrary, $n_{_{\rm S}}\simeq 1$ but $n_{_{\rm
    S}}\neq 1$ is a prediction of inflation. This prediction has not
yet be confirmed by the observations but there is good evidence that
it is true. For instance, the WMAP data indicate that $1-n_{_{\rm
    S}}=0.018^{+0.019}_{-0.02}$~\cite{Larson:2010gs,Komatsu:2010fb,Martin:2006rs}). In
a few months, with the data of the Planck satellite, this prediction
will be definitively confirmed or ruled out with certainty (\ie at the
$5\sigma $ confidence level). Of course, the spectral index is not the
only quantity of interest and a list and a discussion of all the
inflationary predictions can be found in
Refs.~\cite{Martin:2003bt,Martin:2004um,Martin:2007bw}.

\section{Conclusions}
\label{sec:conclusions}

In this short letter, we have discussed the quantum-mechanical aspect
of inflationary fluctuations. In particular, we have argued that
cosmological fluctuations are placed in a very peculiar quantum state,
namely a strongly squeezed state. In the very same way that the CMB
radiation represents the most accurate black body radiation ever
produced (recall that, in the laboratory, it is not possible to
produced such an accurate black body), the squeezed state of the
inflationary perturbations probably represents the strongest squeezed
state ever observed in Nature. As a consequence, one could be led to
conclude that the quantum properties of the CMB anisotropies are
highly non-classical. And, in a sense, they are! But, we have also
shown that everything turns out to be equivalent to a classical
stochastic process. Therefore, even if the statistical nature of the
fluctuations is here to remind us about their quantum origin, all the
genuine quantum-mechanical imprints (such as interferences or the
off-diagonal terms in the density matrix) are ``washed out'' by the
strong squeezing and by decoherence. This explains why astronomers can
forget about the quantum origin of the CMB fluctuations and consider
that they are simply described by a Gaussian stochastic process.

\par

There remains, however, an open question, namely the issue of definite
outcomes. This issue is of course already present in conventional
quantum mechanics (decoherence does not solve the definite outcome
problem, see Ref.~\cite{2012arXiv1208.0904C}) but it is clearly more
embarrassing in the context of cosmology where no observer was present
during inflation~\cite{Sudarsky:2009za}. This problem was recently
discussed in Ref.~\cite{Martin:2012pe}. This is usually taken as an
indication that the Copenhagen interpretation is not well suited to
cosmology and that other theoretical frameworks, such as (for
instance) the many worlds interpretation or the collapse theories,
should be considered. This also shows that inflation is not only a
scenario which allows us to convincingly reproduce various
astrophysical observations but is also a playground where very
fundamental questions can be discussed and studied. Usually this is
considered to be a feature of quantum cosmology only but, in fact,
this is also the case for inflation. It is clear that quantum effects
in the context of inflation are treated in a perfectly standard way
and, moreover, that we have at our disposal detailed observations that
directly originate from those quantum effects. For this reason, it
does not seem necessary to go as far as quantum cosmology (although it
is of course interesting) to discuss very fundamental questions at the
crossroads of quantum mechanics and gravity: inflation can do the job!

\section*{Acknowledgments}

I would like to thank the organizers of the COSGRAV12 conference for
their hospitality. I thank Marc Lilley and Vincent Vennin for careful
reading of the manuscript.

\section*{References}
\bibliography{biblio_jerome_martin}

\end{document}